# Title: A possible alternative evaluation method for the non-use and nonmarket values of ecosystem services

**Authors:** Shuyao Wu[1], Shuangcheng Li[1*]

**Authors affiliations:**

[1] Key Laboratory for Earth Surface Processes of the Ministry of Education Center of Land Science, College of Urban and Environmental Sciences, Peking University, Beijing, China

**\* Corresponding to:** E-mail: wushuyaoalbert@pku.edu.cn

¶These authors contributed equally to this work.

# Abstract

Monetization of the non-use and nonmarket values of ecosystem services is important especially in the areas of environmental cost-benefit analysis, management and environmental impact assessment. However, the reliability of valuation estimations has been criticized due to the biases that associated with methods like the popular contingent valuation method (CVM). In order to provide alternative valuation results for comparison purpose, we proposed the possibility of using a method that incorporates fact-based costs and contingent preferences for evaluating non-use and nonmarket values, which we referred to as value allotment method (VAM). In this paper, we discussed the economic principles of VAM, introduced the performing procedure, analyzed assumptions and potential biases that associated with the method and compared VAM with CVM through a case study in Guangzhou, China. The case study showed that the VAM gave more conservative estimates than the CVM, which could be a merit since CVM often generates overestimated values. We believe that this method can be used at least as a referential alternative to CVM and might be particularly useful in assessing the non-use and nonmarket values of ecosystem services from human-invested ecosystems, such as restored ecosystems, man-made parks and croplands.

# Introduction

Nature provides the foundation of human existence and therefore contains a tremendous value. People perceive these values through ecosystem services, which refer to the benefits people obtained from nature [1]. Indeed, there have been many studies applied the concept of ecosystem services in estimating the use and non-use values of natural capitals. For example, Costanza et al. [2] estimated the economic value of global natural capital by using the values of 17 ecosystem services; Curtis [3] calculated the value of World Heritage Area in Australia by using 20 ecosystem services and goods; Remme et al. [4] used seven ecosystem services values to evaluate the different worth of land in the Limburg province, the Netherlands.

Since not all values from ecosystems are directly perceived by people, the total values of nature can be further divided into use and non-use values [5]. Use values can be further categorized into direct use value, indirect use value and option value. Non-use values are mainly composed of existence and bequest values [6,7]. Although every ecosystem service possesses both use and non-use values, most provisioning services (e.g. food, water and fuel supply) and regulating services (e.g. climate regulation, waste treatment, soil retention, etc.) can be directly or indirectly utilized by people. This makes these services usually contain more use-values. On the other hand, the values of many cultural and some regulating services, such as aesthetic enjoyment, cultural information and biodiversity maintenance, are non-use or nonmarket in many cases [8,9].

Despite the disputation on the validity of monetizing the values of ecosystem services, monetization of these values is at least necessary especially in the areas of policy decision-making, environmental cost-benefit analysis and environmental impact assessment [10,11]. Several valuation methods have been developed to perform this task. For values that can be directly reflected by commercial market prices, demand-based valuation methods such as market price, travel cost and hedonic pricing can be applied [12,13]. If there is no existing market that can be used as a direct reference for prices, supply-based methods, such as production function, and cost-based methods, such as replacement cost and avoided damage cost, can be used to simulate market conditions for valuation [12,13]. Moreover, for values that cannot be evaluated in the conventional market economy, like non-use values and option value, choice experiment and contingent valuation methods are options for appraisal.

The contingent valuation method (CVM), has become one of the most widely applied valuation methods for non-use and nonmarket values [10,14]. This method basically asks respondents to provide an estimate of how much money they would be willing to pay (WTP) for a certain good or service in a hypothetical or contingent market and then deems this WTP as the value of the good or service [15]. However, its reliability has been long criticized because of the biases that associated with this method, such as embedding effect, sequencing effect, payment vehicle bias, information effect, elicitation effects, hypothetical bias, strategic bias, yes-saying bias etc. [10,16]. Although researchers have been trying to minimize the effects of these biases, studies still found CVM could generate inaccurate estimations [17-19].

In order to provide some alternative valuation results for comparison purpose, here we proposed the possibility of using an alternative method that incorporates fact-based costs and preferences for evaluating non-use and nonmarket values, which we referred as value allotment method (VAM). In this paper, we discussed the economic principles of VAM, introduced the performing procedure, analyzed assumptions and potential biases that associated with the method and compared VAM with CVM through a case study for the reliability test.

# Value allotment method (VAM)
## Economic foundation

In essence, the VAM asks people to allot the weight of components of an asset's total value, in order to appraise the components with non-use and nonmarket values. In our case, the asset refers to the ecosystem that provides ecosystem services and the components are the services or good with instrumental functions or just intrinsic values. This method is viable under the assumption that the total value of an asset is the sum of all of its component values. In turn, the value of any particular component can also be determined by the total value and the proportion of this component weights in the total value. The proportion of a component weights in the total value can be obtained by stated preferences from people through surveys or interviews. Therefore, the general equation for valuating the components with non-use or nonmarket values is the following:

$$Component\ Value\ =\ Total\ Asset\ Value\ *\ Value\ Component\ Allotment\ *\ Contingent\ Adjustment \tag{1}$$

In conventional asset appraisal world, there are three generally accepted valuation approaches for asset appraisal: cost, income and market approaches. The income approach estimates the value of an asset by aggregate and discount the discrete forecast of future benefits to present at an appropriate discount rate [20]. The market approach evaluates the value of an asset by comparing it with a similar subject or asset that has recently transaction in a market [20]. Since not all environmental assets have comparable tradable markets or incomes that can represent the total value but many can be paid to reproduce. For example, there is no market for trading the whole Amazon forest and a small patch of stand in a pristine forest doesn't generate direct market values. We decided to use the cost approach to estimate the total value of environmental assets. The cost approach relies on the economic principle of substitution, which utilizes the amount of money required to replace the service capability of an asset to quantify the value of it [21].

## Value determination

The total value determination section contains three major equations as follows:

*Total Asset Value = Replacement Cost – Physical Deterioration – Incurable Functional Obsolescence – Economic Obsolescence* (2.1)

*Replacement Cost = Reproduction Cost – Curable Functional Obsolescence* (2.2)

*Reproduction Cost = Direct Cost + Indirect Cost* (2.3)

As shown in equation 2.1, there are three types of adjustments, namely, physical deterioration, functional obsolescence and economic obsolescence [21]. The definition and examples in an environmental asset (e.g. a patch of forest) of each term can be found in Table 1.

**Table 1. The definition and example of each term that used in cost appraisal approach for environmental assets. Take a forest patch as an example.**

| Terms | Definition | Examples |
|---|---|---|
| **Reproduction Cost** | The total cost, in current prices, to develop a new exact duplicate of the asset | The cost of using all identical species with same structure and same configuration to recreate a new forest |
| **Replacement Cost** | The total cost, in current prices, to develop a new asset having the same functionality | The cost of using species with better functional performance and lower market prices to recreate a forest |
| **Direct Cost** | The cost, in current prices, that can be completely attributed to the production of specific goods or services | The market prices of plants, soil, fertilizer and the cost of labor, equipment and shipping that used during construction |
| **Indirect Cost** | The cost, in current prices, that cannot be completely attributed to the production of specific goods or services | The cost of maintenance after construction and legal fees |
| **Physical Deterioration** | The reduction in value because of physical wear and tear | Leaf or tree death that caused by disease, insects or aging |
| **Curable Functional** | The reduction in value due to inability to perform the function | Retarded tree growth rate due to the usage of inappropriate |
| **Obsolescence** | for which the asset was originally designed. However, this obsolescence can be cured through substitution or addition, given the cost of replacing it must be the same as or less than the expected increase in value | non-local species and it is cost-effective to use local species |
| **Incurable Functional Obsolescence** | The reduction in value due to inability to perform the function for which the asset was originally designed. However, this obsolescence cannot be practically or economically corrected and may be caused by deficiencies or superadequancies | Retarded tree growth due to severe competition, which can be caused by too high plant density |
| **Economic Obsolescence** | The reduction in value due to the locations, events or conditions that are external to the current use or condition of the asset | Retarded tree growth due to unsuitable climate |
| **Remaining Useful Life** | The number of remaining years of an asset to be able to function in accordance with its intended purpose | The weighted average remaining life of the dominant type of vegetation (e.g. every grass in a grassland or every tree in a forest) |

Remaining useful life (RUL) is another factor that needs to be considered in the cost determination process [21]. In environmental and natural asset valuation, RUL can give an indication of the maximum validity period of valuation results. The estimated value within RUL can also be adjusted yearly according to certain functions, such as inflation rate. For traditional business valuation, there are many factors can influence the useful life of an asset, such as legal, contractual, functional, technological, economic and analytical [21]. However, the functional factor might be the only applicable one in environmental asset RUL determination. Therefore, the remaining life of an environmental asset should be the number of remaining years of an asset to be able to function in accordance with its intended purpose. Since many environmental assets don't possess a conventional function expiration date because of their self-regeneration ability, we can only estimate the remaining useful life of an environmental asset excluding the effects of future regeneration in this case. We need to only focus on the asset under valuation at the moment and their changes from the moment to the future. For instance, we can calculate the weighted average remaining life of all trees in a forest and leave out the future regenerated tree to obtain an RUL estimation of the forest currently under valuation.

In order to ensure a generally acceptable accuracy level, there are four assumptions people need to be aware before applying the cost valuation approach [22]. The first assumption is that the asset under valuation is and will still be used. This assumption would be met unless the environmental asset, such as a patch of forest, under valuation is severally devastated. The second assumption is that there is relatively adequate information about the asset in order to generate accurate results. In order to meet this assumption, people need to collect data about the environmental asset beforehand. If we still use a patch of forest as an example, data such as species composition, number of species, height and weight, growth condition, foliage health etc. are necessary. The third assumption is that the asset under valuation is reproducible. Most environmental assets should be able to meet this assumption. For instance, a forest can be replanted, and a river can also be reconstructed in many cases. The last assumption is that the value of the asset under valuation is depreciating. This assumption will be met in plant-dominated ecosystems if the self-reproduction ability of vegetation is excluded from the scope of the appraisal. In developing ecosystems, such as a young forest stand, the cost valuation approach would usually generate underestimated total value. However, if we are not interested in the maximum but just the current total value of an ecosystem, we can still use the cost approach to estimate the total value of a young ecosystem at the moment of appraisal. Furthermore, adjustments, such as inflation rate or ecosystem development forecast model, could also be applied to obtain the maximum total asset value of young ecosystems.

## Allotment determination

The next step of VAM is acquiring hypothetical appraisers' preferences on the weight allotment of different components of the asset's total value through interviews and/or questionnaires. At this stage, various question forms can be used to acquire the most important information from people: how much do they think each component of the asset worth? As aforementioned, ecosystem services can be used as a useful tool for conceptualizing different value components of an environmental asset [23]. Provisioning, regulating and cultural services that an environmental asset provides can be viewed as the components of the asset value. Supporting services, such as nutrient cycling and soil formation, should be excluded to avoid double counting because the values of these services are usually reflected in other service types [24].

Additionally, there are two more major problems need to be considered in the allotment determination stage in order to obtain accurate results. Firstly, how to ensure the accuracy and completeness of the value components? Secondly, how to ensure the accuracy of the allotment decision? If these two problems were not carefully dealt with, the obtained allotment might be biased. Providing adequate information to interviewees should be able to effectively alleviate the first problem. All relevant information such as asset type, purpose, location, size, composition, current conditions, surrounding environments etc. should be provided to interviewees in both textual and graphic formats before allotment determination.

Even with sufficient information provided, people might still not be able to come up with a complete and accurate candidate list for value components due to reasons such as inexperience, misunderstanding and preconceptions. Since a high-quality list is crucial for estimation accuracy, we believe that expert knowledge is required to provide guidelines for common interviewees during this process. For example, ecosystem services researchers can create a list of some most important ecosystem services that the asset under valuation provides based on thorough investigations. Interviewees then can use the list as a reference. However, respondents should still be asked explicitly for their own ideas on asset value components.

In terms of the second problem, utilizing a proper elicitation method should be able to improve the allotment accuracy effectively. In CVM, there are four common types of elicitation methods, which are the bidding game, payment card, open-ended and dichotomous choice approach [25]. All of them can be also utilized in VAM as long as the question designers ensure the sum of each component allotment equals 100%. This premise might make the open-ended approach a more preferred option due to its simplicity. For example, a table that contains all value components and space for filling allotment for each component can be provided to interviewees (see

S1 for examples). It is important to keep in mind that the open-ended method might not work very well in situations where the respondents are not familiar with the asset under valuation. Venkatachalam [10] in his review of CVM concluded that each elicitation method has its pros and cons. Therefore, elicitation method selection in a VAM study more relies on factors such as the nature of the asset investigated, study budget, nature of the respondents targeted, nature of the statistical technique used, etc..

Once we multiply the total asset value with the value allotment of each component, we will obtain a rough estimation of the value of each component, including those with non-passive values. However, apart from the physical deterioration, functional obsolescence and economic obsolescence for value adjustment, we also suggest adding an adjustment coefficient, which should also obtain from respondents. Since most environmental assets are not products that can be traded in conventional markets, we believe that this adjustment coefficient is necessary to further decrease any potential hypothetical bias. The adjustment coefficient measures the opinion of hypothetical appraisers on the proper component value of an asset after they obtained the estimation based on their allotment and cost-based estimated value. For example, if an interviewee believes that the estimated aesthetic value of the forest under review is 10% lower than the calculated allotment value, the contingent adjustment then will be 110%. Figure 1 shows a flowchart that summarizes the key steps of VAM.

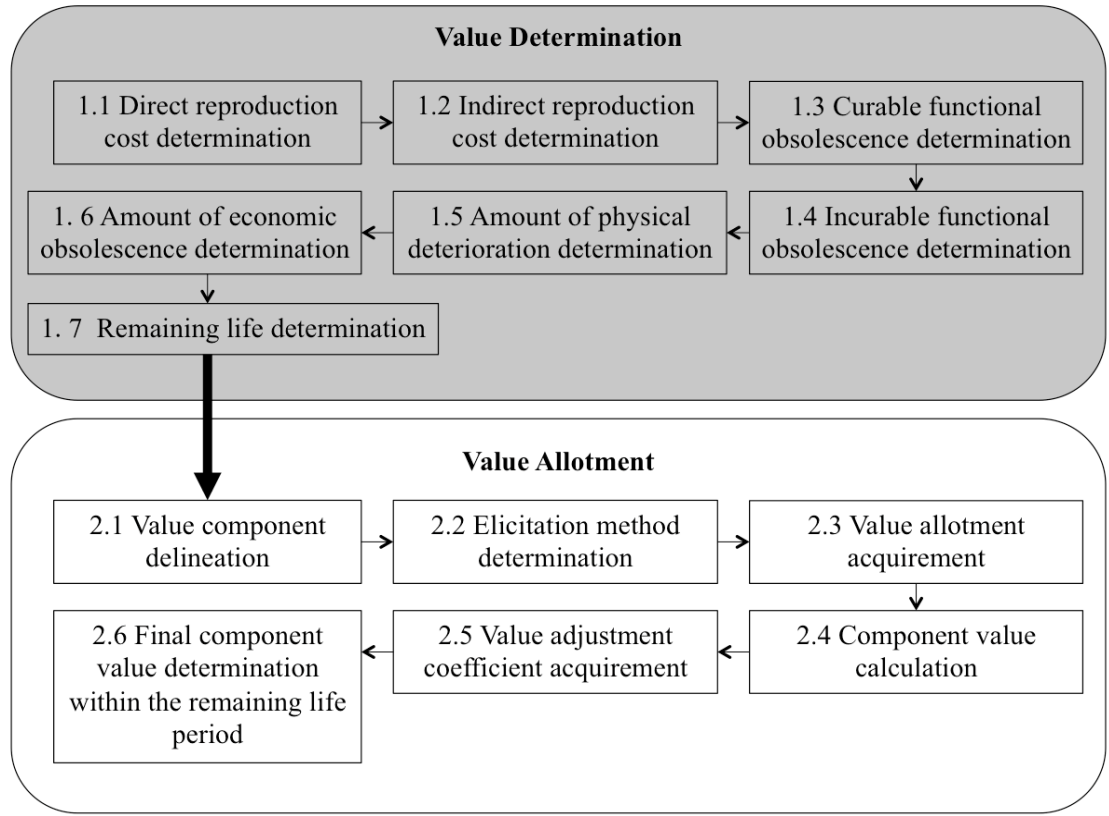

**Figure 1.** Flowchart of the Value Allotment Method.

# VAM Advantages, Bias and Potential Solutions

Although both VAM and CVM are mainly designed to evaluate non-use and nonmarket values, compare to the CVM, VAM might be particularly useful in assessing the non-use and nonmarket values of manmade environmental assets with real investments, such as restored ecosystems, man-made parks and croplands. For these assets, a few interviews with the real investors might be enough to effectively reflect more realistic figures of the non-use values rather than surveying a large number of hypothetical payers.

In addition, VAM and CVM are also very different in their evaluation approaches and validation period. Unlike the cost approach based VAM, CVM in environmental value assessment took a market approach since it essentially created a hypothetical market to acquire a comparable consumer surplus measure of nonmarket public goods, such as ecosystem services [13,20]. Furthermore, the validation period of VAM and CVM are also different. The result validation period of VAM results is determined by the calculated remaining useful life of the asset under appraisal. On the other hand, the result validation period of CVM results is determined by variable opinions from interviewees.

In terms of method validity, we believe that the VAM should be able to deal with at least two types of bias better, which are the hypothetical and embedding bias. Since VAM provide a more objective foundation for evaluation, the complete hypothetical conditions in CVM should be alleviated. Moreover, the clear reference value that VAM provides also eradicates potential confusions respondents may have regarding appraisal boundary. A detailed comparison of some key aspects between VAM and CVM can be found in Table 2.

**Table 2. Comparison of some key aspects between contingent valuation method and value allotment method in ecosystem service valuation.**

|  | **Contingent Valuation Method** | **Value Allotment Method** |
|---|---|---|
| **Target** | Non-use and nonmarket values | Non-use and nonmarket values, more suitable for assets with real investments |
| **Approach** | Market-based approach | Cost-based approach |
| **Basic Procedures** | Survey to acquire willing to pay | Cost-based value determination + Survey to acquire value allotment + Survey to acquire adjustment coefficient |
| **Validation period** | Validation period determined by interviewees' opinion | Validation period determined by remaining useful life |
| **Reliability: Hypothetical Bias** | Since the CVM uses completely hypothetical scenarios, there is a potential divergence between the real and hypothetical payments. | VAM provides a clear and more objective foundation for non-use and nonmarket value evaluation. |
| **Reliability: Embedding Bias** | Since CVM can't differentiate a clear foundation for people, people's willingness to pay for the same good can be depending on whether the good is valued on its own or | Since VAM delineates a distinct value boundary for people, people can use this relatively objective value as a clear reference point. |

| | valued as a part of a more inclusive package [26]. | |

Despite the above-mentioned merits that VAM may have, there are also many potential uncertainties and biases associated with the method. There are two major categories of error, one relates to the cost determination and another relates to the value allotment. The first and foremost question we need to ask about VAM is can the cost approach reflect the true total value of an asset. In terms of the value allotment, some effects may affect its accuracy include starting point and sequence bias, strategic bias and information bias. More detailed summary of each bias type can be found in Table 3.

**Table 3. The descriptions and potential solutions for the bias types of the value allotment method.**

| Bias Types | Description | Potential Solutions |
|---|---|---|
| **Cost Approach Problem** | Cost approach might not be able to reflect the true value of an asset | Only use VAM to appraise non-use and nonmarket values that are unsuitable for market and income approaches |
| **Value Estimation Error** | Cost of replacement or reproduction might not be fully or accurately monetized | Ensure adequate information collected and all costs and obsolesces are considered beforehand |
| **Starting Point and Sequence Biases** | The starting point and sequence of options that used in elicitation methods might influence the final value allotment | Use pre-survey to determine suitable range for starting point; State explicitly that the sequence and starting point of the value components are not implying any form of value-importance relationship |
| **Strategic Bias** | There might be free riding or overpledging people who understate or overstate, prospectively, their allotment | Ensure acceptable data size (at least >30 responds) and exclude extreme values during analysis |

| | of value components because they try to use their decisions to make actual changes. | |
|---|---|---|
| **Information Bias** | The nature and amount of information provided beforehand might influence the final value allotments | Use pre-surveys to determine how much information should be presented; Ensure respondents understand all relevant and accurate information in both textual and graphic formats |

# Case study

We used the National Baiyun Mountain Scenic Area (BMSA) in Guangzhou, China as our case study for testing the criterion validity of VAM, which may refer to the correspondence between VAM results and a criterion results such as CVM in this case [27]. Chen and Jim [28] used the contingent valuation method (double-bounded dichotomous choice as the elicitation method; annual conservation tax lasting 5 years as the payment vehicle) to acquire the WTP for the total ecological value in BMSA from 562 interviews and found that the medium WTP was 149 RMB (67-228 at 95% confidence interval) per household per year. According to the method, the aggregated total value of BMSA in 2007 was estimated to be 291 million RMB (131-446 at 95% confidence interval; approximately 38.2 million US$) per year. Based on acquired people's motivation for urban biodiversity conservation and the relative importance of floral species, the value of floral diversity worth approximately 15.6 million RMB. We applied the VAM method to estimate the floral diversity value in the same area as for comparison purpose.

## Total value of BMSA

The National Baiyun Mountain Scenic Area (BMSA) is located in Guangzhou, China (113°16'~113°19'E and 23°09'~13°13'N) (Fig 2). It has sub-tropical monsoon climate [29]. The average annual temperature and precipitation are 21.7°C (13.2°C ~28.5°C) and 1727.4mm [29]. The total coverage area is approximately 20.8 km$^2$ [28]. According to the latest biotic survey, there are 876 vascular plant species in total [30]. The plant community can be divided into five vegetation types in BMSA, which are evergreen coniferous forest (~2.69 km$^2$), evergreen mixed forest (~4.32 km$^2$), evergreen broadleaved forest (~12.72 km$^2$), savanna (~5.2 km$^2$) and orchard (~0.11 km$^2$). The total vegetation coverage is about 95%. [31]. The BMSA was almost deforested before 1951 and reforested with *Pinus massoniana* during 1951 to 1953. During 1995 to 1999, the forest stand is replanted with various broadleaf species in order to increase biodiversity [32]. The current average tree and shrub density are approximately 1085 and 1596 per km$^2$ [33-36].

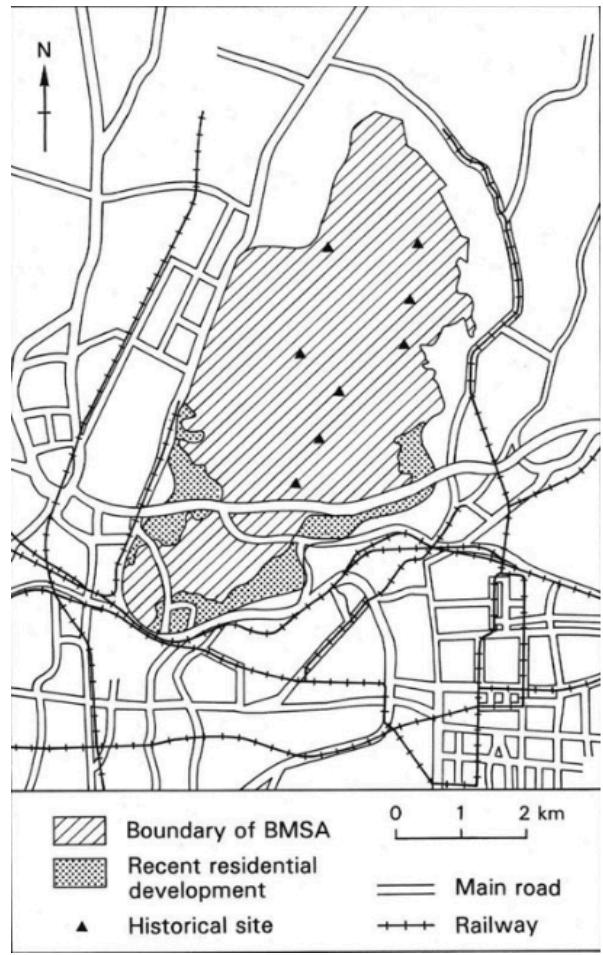

**Figure 2. Map of Baiyun Mountain Scenic Area (BMSA) in Guangzhou, China. Adapted from Chen and Jim (2010).**

In addition, Liang and Li [37] found that about 9.8% of BMSA area is invaded by 5 invasive species, which are *Mikania micrantha*, *Ipomoea cairica*, *Wedelia trilobata*, *Lantana montevidensis* and *Bougainvillea spectabilis*. Insects also caused approximately 3.32% mortality in *Pinus massoniana* stands across BMSA [37]. Zeng et al. [34] and Jia et al. [35] tested the annual growth rate and carbon sequestration ability between common non-local tree species and local tree species and found that the non-local species performed better in both categories.

Based on the above-mentioned information, it is estimated that there are approximately 2.15 million trees and 2.97 million shrubs in BMSA. Due to the relatively low tree and shrub density, grass species are present throughout BMSA except in orchard and covers an area of approximately 2,493 $hm^2$. According to the *Budget Making Standards for Land Development and Consolidation Projects* by Ministry of Finance and Ministry of Land and Resources of P.R. China [38] and 2007 discount rate, the average establishing cost (includes seedling cost, equipment, labor and other costs such as legal and administration) for tree, shrub and grass are 86.0 RMB per tree and shrub seedlings and 36,000 per $hm^2$ for grass seeds in 2007 value. The annual maintenance fee is roughly 26.9 RMB per tree and 3.6 RMB per $hm^2$ for shrub and grass species [39,40]. Till the year of 2007, the required maintenance year for *Pinus massoniana* is about 53 years and 11 years for other plant species.

Consequently, we obtained a replacement cost of 1.60 billion RMB (approximately 210 million US$ by Chan and Jim [28]'s exchange rate) for the plant community in BMSA by adding a direct cost of 605.9 million RMB (219.3, 298.8 and 87.8 million RMB for tree, shrub and grass, respectively) and indirect cost of 996.2 million RMB (993.1, 1.20 and 1.86 million RMB for tree, shrub and grass, respectively) together (Table 4). All the area that covered by invasive species (~2 $km^2$) was replaced by non-invasive species in the replacement in order to eliminate curable functional obsolescence. Insect-caused mortality was responsible for approximately 14.3 million RMB, which should be regarded as physical deterioration. Although the plant density was lower than nearby reference forest ecosystems, there was no sign of any negative effect on growth or survival rate in literature [41], which suggested that there was no obvious incurable functional obsolescence. In terms of economic obsolescence, since the majority of the plants were suitable in this environment, we deemed that there was no significant economic obsolescence [34,35]. Therefore, the total value of biodiversity in BMSA was estimated to be 158.8 million RMB in 2007 value.

**Table 4. The 2007 value of the variables and sub-variables of total value determinants in Baiyun Mountain Scenic Area (BMSA).**

| Variables | Sub-variables | Values (Unit) |
|---|---|---|
| Replacement Cost | Direct Cost (plants and labor): | 605,932,981 RMB |
|                              | Indirect Cost (maintenance and other):   | 996,163,855 RMB                          |
|                              | Total Cost                               | 1,602,096,836 RMB (~209,989,519 US$)     |
| Physical Deterioration       | Tree infection rates by diseases and Insects | 14,309,910 RMB                       |
| Incurable Functional Obsolescence | Species richness deficiency         | 0 RMB                                    |
| Economic Obsolescence        | Percentage of unsuitable species         | 0 RMB                                    |
| Total Asset Value            | Replacement cost minus all deprecation factors | 1,587,786,926 RMB                  |
| Remaining Useful Life        | Weighted average life                    | 64 years                                 |

## Value allotment of BMSA

The questionnaire that we used in this study comprised three sections referred to as A, B and C. Section A firstly introduced the purpose of the survey and then obtained some basic demographic information about respondents, such as gender, age, education level and income level. They were also asked whether they had been to and lived in Guangzhou, the frequency of visiting natural ecosystem and frequency of visiting BMSA. Section B presented some essential information about BMSA, such as location, area size, climate, species composition, ecological condition, environmental quality etc., in both text and graphic formats to respondents. In Section C, respondents were firstly given the total ecological value of BMSA based on floral species replacement cost (~1.721 billion RMB) and some possible value component options with explanations, which are carbon sequestration and oxygen generation, water yield, soil retention, biodiversity maintenance, microclimate regulation, recreation, aesthetic enjoyment and air purification [42,43]. They were also asked to add any other component that they deem suitable for BMSA in this section. Then they were asked to allot the weight of different ecosystem services components of BMSA's total ecological value in percentage with an open-ended question form (Table 5). Based on their % allotment on "biodiversity maintenance", they were given a chance to adjust the estimated component value by adding or subtracting certain percent of the estimated value (Table 5). A total of 369 questionnaires were distributed electronically. The questionnaires were not only targeted to people who lived in Guangzhou since we believe that most people should be able to provide their opinions on the value component allotment if given enough information. Since median is more robust and less sensitive to the specification of the distribution function, we used the median opinions instead of the mean to represent the allotment of each component.

**Table 5. Sample questionnaire questions that used for acquiring value allotment and contingent adjustment acquirement in Baiyun Mountain Scenic Area (BMSA) case study.**

| Based on the provided information, please allot the weight of different ecosystem services components of BMSA's total ecological value. |  |
|---|---|
| **Note**: The sequence of the component is completely random and not implicit. If you don't think the ecosystem service is a part of value component, please write 0%. The total sum of all allotment should be 100%. | |
| **Possible Value Component** | **Value Allotment (%)** |
| Carbon Sequestration and Oxygen generation | |
| Water Yield | |
| Soil Retention | |
| Biodiversity Maintenance | |
| Microclimate Regulation | |
| Recreation | |
| Aesthetic Enjoyment | |
| Air Purification | |
| Others | |
| Based on your allotment (X%), the <u>Biodiversity Maintenance</u> component would have a value of "<u>allotment-calculated</u>". According to your opinion on biodiversity, do you think this value is an appropriate estimation for this component? | |
| ☐ No, it is underestimated by about ________ %. <br> ☐ Yes, it is about right. <br> ☐ No, it is overestimated by about ________ %. | |

## Results and discussion

There were a total of 120 responses acquired. Three of them contained invalid or missing answers, which made 117 qualified for further analysis. The medium allotment and estimated value of biodiversity maintenance in BMSA were 10% and 158.8 million RMB in 2007 value (still 158.8 at 95% confidence interval according to bootstrap analysis), respectively. After allotting their preferences on value component, there were 21 out of 117 (17.9%) respondents reported an underestimation of the biodiversity maintenance value and increased the estimation by 205% on average. However, there were also 5 out of 117 (4.3%) respondents reported an overestimation and decreased the value by 39% on average. The rest of respondents (91 out of 117, 77.8%) didn't provide any adjustment coefficient.

Since the annual biodiversity value in BMSA that obtained from Chen and Jim [28] need to be accumulated to compare with the VAM results, we used the following equation (3) to acquire the total present value of biodiversity in BMSA:

$$C = R + \frac{R}{(1+r)} + \frac{R}{(1+r)^2} + \cdots + \frac{R}{(1+r)^{(T-1)}} \quad (3)$$

where *C* is the accumulated present value of biodiversity of BMSA; *R* is the annual biodiversity WTP value; *r* is discount rate; *T* is the period of accumulation. We used the 2007 national average deposit interest 3.46% as the *r* factor and the remaining useful life 64 years as the *T* factor for calculation. As the result, the total value from CVM was found to be approximately 412.6 million RMB (193-639 at 95% confidence interval), which is roughly 2.6 times higher than the VAM value. We were unable to verify the statistical significance of the difference due to the lack of CV data. However, we believe that the difference should be meaningful since there is no slight overlap even between the two 95% confidence intervals.

The reasons for this disparity can be explained from two perspectives. On the one hand, many studies have shown that the WTP from CVM method can be overstated because of hypothetical bias, information effects and flawed experimental design [44-46]. For instance, Neill et al. [47] found that the hypothetical WTPs of a watercolor painting and a framed 16$^{th}$-century world map were both significantly higher than actual WTPs. Foster et al. [19] also drew the same conclusion after comparing the actual donations to environmental preservation in the UK and the hypothetical WTP values from six CV studies.

On the other hand, our cost-based estimated total ecological value of BMSA might also be too conservative due to several reasons. Firstly, the "Budget Making Standards Book for Land Development and Consolidation Projects" by the Ministry of Finance and Ministry of Land and Resources of China, which is the reference for our plants prices, doesn't differentiate the prices among species but just gives averages for trees, shrubs and grasses. Secondly, we didn't consider the shipping fee of plants that might occur during replacement cost determination because there is no reliable source for such estimation. This could potentially lead to a significant underestimation to the total replacement cost. Thirdly, we only estimated the replacement cost of the floral ecosystems in BMSA since we believe that they are the main provider of ecosystem services thus the main value holder. However, if we include the replacement cost of faunal species and different landscape features such as lakes and streams in BMSA, the estimated total value would be higher. Last but not least, we used the open-ended elicitation approach for acquiring contingent adjustment coefficient due to its convenience. Nevertheless, this technique tends to give conservative estimates than elicitation approaches such as the bidding game [48]. Since more people believe that the estimated biodiversity maintenance value is an underestimation instead of overestimation (21 versus 5), the value could be larger if other elicitation technique was applied.

## Conclusions

In this paper, we designed a method, which referred as the value allotment method, to evaluate the monetary value of environmental assets. In principle, VAM asks people to allot the weight of components of an asset's total value, in order to appraise those components with non-use and nonmarket values. In other words, if one environmental asset needs to be paid to rebuild, how important is each reason contribute to this replacement cost. We use this importance contribution of each reason, especially those reasons with non-use and nonmarket values to the total replacement amount as the surrogate of this reason's value.

The heuristic example of BMSA in Guangzhou, China showed that the VAM gave more conservative estimates than the CVM, which could be a merit since CVM often generates overestimated values [10]. We believe that this method can be used at least as a referential alternative to CVM and might be particularly useful in assessing the non-use and nonmarket values of man-made environmental assets with real investments. We also propose with the caution that it is possible to extend its application to more conventional asset non-use and nonmarket value appraisal. Last but not least, future researches that focus on potential biases alleviation are also suggested for better result reliability.